\documentstyle [a4,12pt]{article}
\newcommand{\mincir}{\raise -2.truept\hbox{\rlap{\hbox{$\sim$}}\raise5.truept
\hbox{$<$}\ }}
\newcommand{\magcir}{\raise -2.truept\hbox{\rlap{\hbox{$\sim$}}\raise5.truept
\hbox{$>$}\ }}
\newcommand{\minmag}{\raise-2.truept\hbox{\rlap{\hbox{$<$}}\raise 6.truept\hbox
{$>$}\ }}
\newcommand{\be}{\begin{equation}}
\newcommand{\ee}{\end{equation}}
\newcommand{\ba}{\begin{eqnarray}}
\newcommand{\ea}{\end{eqnarray}}
\newcommand{\brr}{\begin{array}}
\newcommand{\err}{\end{array}}
\newcommand{\bc}{\begin{center}}
\newcommand{\ec}{\end{center}}

\newcommand{\et}{et al.~}

\title{ {\bf TESTING THE VELOCITY FIELD IN NON--SCALE INVARIANT COLD
DARK MATTER MODELS}}
\author{
{\bf Lauro Moscardini}$^1$, {\bf Giuseppe Tormen}$^{2,1}$, \\
{\bf Sabino Matarrese}$^3$ \& {\bf Francesco Lucchin}$^1$ \\ ~\\
{\em $^1$Dipartimento di Astronomia, Universit\`a di Padova,} \\
{\em vicolo dell'Osservatorio 5, I--35122 Padova, Italy} \\ ~\\
{\em $^2$Institute of Astronomy, University of Cambridge,}\\
{\em Madingley Road, Cambridge CB3 0HA, UK} \\ ~\\
{\em $^3$Dipartimento di Fisica G. Galilei, Universit\`a di Padova,} \\
{\em via Marzolo 8, I--35131 Padova, Italy} \\ ~\\}

\date{}

\begin{document}

\maketitle

\newpage
\section*{Abstract}
We analyze the cosmic peculiar velocity field as traced by a sample of 1184
spiral, elliptical and S0 galaxies, grouped in 704 objects. We perform a
statistical analysis, by calculating the bulk flow, Cosmic Mach Number
and velocity correlation function for this sample and for mock catalogs
extracted from a set of N--body simulations. We run four cold dark matter
(CDM) simulations:
two tilted models (with spectral index $n=0.6$ and $n=0.8$), the
standard model ($n=1$) and a ``blue" one ($n=1.2$), with different values of
the linear bias parameter $b$. By means of a Maximum Likelihood analysis
we estimate the ability of our models to fit the observations, as measured by
the above statistics, and to reproduce the Local Group properties. On the
basis of this analysis we conclude that the best model is the unbiased
standard model $(n,~b)=(1,~1)$, even though the overall flatness of the joint
likelihood function implies that one cannot strongly discriminate
models in the range $0.8 \le n \le 1$, and $1 \le b \le 1.5$.
Models with $b \geq 2.5$ are rejected at the $95\%$ confidence level.
For $n=0.8$ the values of $b$ preferred by the present analysis,
together with the {\em COBE} data, require a negligible contribution to
$\Delta T/T$ by gravitational waves.
Finally, the blue model, normalized to {\em COBE}, does not provide
a good fit to the velocity data.

\vspace{0.5cm}
\noindent{\em Subject headings:} cosmology: theory -- dark matter -- galaxies:
formation, clustering -- large-scale structure of the Universe -- early
Universe.
\vspace{0.5cm}

\section {Introduction}

It is well known that the standard cold dark matter (hereafter SCDM) scenario
for structure formation possesses a high predictive power in explaining many
observed properties of the large--scale structure of the universe. The model is
characterized by a primordial scale--invariant power--spectrum, $P(k) \propto
k^n$, with spectral index $n=1$, of Gaussian adiabatic perturbations in an
Einstein--de Sitter universe with vanishing cosmological constant. As usual,
one can parametrize the amplitude of the primordial perturbations by the
linear {\em bias} parameter $b$, defined as the inverse of the {\rm rms} mass
fluctuation on a sharp--edged sphere of radius $R_8\equiv 8~h^{-1}$ Mpc.
In this paper we adopt the value $h=0.5$ for the Hubble constant $H_0$ in
units of $100$ km ${\rm s^{-1}}$. The {\em COBE} DMR detection
(Smoot \et 1992; Bennett \et 1994) of large angular scale Cosmic Microwave
Background (CMB) anisotropies can be used
to normalize the SCDM power--spectrum, resulting in $b \approx 0.9$.
However, the SCDM model has met increasing problems, mostly due to the high
ratio of small to large--scale power; in particular the {\em COBE} normalized
model predicts excessive velocity dispersion on Mpc scale and is unable to
reproduce the slope of the galaxy angular correlation function obtained from
the APM survey (Maddox \et 1990).

Many alternative models have been proposed to overcome the difficulties of the
SCDM model. In order of increasing number of changes of the standard
assumptions one can quote: $i)$ ``tilted" (i.e. $n < 1$) CDM models (hereafter
TCDM); $ii)$ hybrid (i.e. hot plus cold dark matter) models; $iii)$ CDM
models with a relic
cosmological constant; $iv)$ CDM models with non--Gaussian initial conditions.

In this paper we will mostly consider TCDM models. This can be considered the
most natural way to decrease the small--scale power relative to the
large--scale one: in fact, tilting the spectral index of the primordial
perturbations boosts power from small to large scales (e.g. Vittorio,
Matarrese, \& Lucchin 1988; Tormen, Lucchin, \& Matarrese 1992; Adams \et 1992;
Cen \et 1992; Tormen \et 1993, hereafter TMLM). Moreover,
these models can be easily motivated in the frame of the inflationary origin
of perturbations (Lucchin \& Matarrese 1985; Adams \et 1992).
The first year data from the {\em COBE} DMR experiment renewed
the interest in these models: the observed anisotropy is in
fact consistent with $n=1.1 \pm 0.5$ on scales $\geq 10^3~h^{-1}$ Mpc.
It is evident
that the COBE normalization of TCDM models implies reduced power on all scales
below $10^3$ Mpc. Moreover, a large number of post--{\em COBE} analyses (see,
e.g., Crittenden \et 1993 and references therein) pointed out that properly
accounting for the gravitational--wave contribution to the Sachs--Wolfe effect
may lead to a relevant modification of the linear biasing factor. One can write
\be
b(n) \approx b_0(n) \times G(n),
\ee
where the quantity $b_0(n)= 0.9 \times 10^{1.2(1-n)} (1 \pm 0.25)$
accounts for the effect of modifying the spectral slope.
The possible contribution to CMB anisotropies by gravitational waves is
taken into account by the function $G(n)$; this effect is especially relevant
in power--law inflation: Lucchin, Matarrese, \& Mollerach (1992) obtained the
approximate relation $G(n) = [(14-12n)/(3-n)]^{1/2}$ in the range
$0.5~\mincir n \leq 1$. Models where the value of $n$ is smaller than unity
but the gravitational--wave contribution is negligible, i.e. $G(n) \approx 1$,
are the typical outcome of ``natural" inflation (e.g. Adams \et 1992).
Even though second year {\em COBE} data have
raised the range of preferred values of the spectral index to
$n = 1.59^{+0.49}_{-0.55}$ (Bennett \et 1994), we can safely continue to use
the above estimate of $b_0(n)$ outside the new range, due to the flatness of
the likelihood function for these observables (e.g. Scaramella \& Vittorio
1993).

For the sake of comparison and completeness, in the present paper we will also
consider an ``anti--tilted" or ``blue" ($n>1$) CDM model (hereafter BCDM).
Recently, many observational data seem to suggest values of $n$ larger than
unity from large--scale CMB anisotropies (Bennett \et 1994; Wright \et 1994;
Hancock \et 1994) and analyses of the matter distribution
(Piran \et 1993; Lauer \& Postman 1994). Contrary to the widespread belief
that inflation always implies $n \leq 1$, blue spectra may be also easily
motivated in the frame of inflation (e.g. Mollerach, Matarrese, \& Lucchin
1994). These models predict a negligible
contribution from gravitational waves to CMB anisotropies, i.e. $G(n) \approx
1$. According to the above equation, BCDM
models, normalized to {\em COBE}, exacerbate the problem of SCDM: too much
power on small and intermediate scales. The simplest solution is to
invoke a large free--streaming effect on these scales, as can be due to a
suitable amount of hot dark matter (Lucchin \et 1994a).

In a recent work (TMLM) we analyzed the peculiar velocity field traced by the
same sample considered here, to probe the primordial spectrum up to scales
$\sim 100~h^{-1}$ Mpc. The results were then compared to similar analyses
carried out on mock catalogs extracted from Monte Carlo simulations. These
were obtained from linear theory in $n \leq 1$, and/or $\Omega_0 \leq 1$ CDM
models, for different values of the bias factor, with the assumption that the
galaxy velocity field gives an unbiased signal of the underlying mass
distribution. In the present paper the same type of analysis is performed on
mock catalogs extracted from N--body simulations in $\Omega_0 = 1$ CDM models.
A similar method has been applied also to simulations with skewed (i.e.
non--Gaussian) CDM initial conditions: the results
will be presented in (Lucchin \et 1994b).

The plan of the paper is as follows. Section 2 is devoted to the description of
the real catalogs of galaxy peculiar velocities and of the method used to
construct the mock catalogs from the N--body simulations. In Section 3
the results of the different statistical tests adopted in the study of
the velocity field are presented and a Maximum Likelihood analysis is
performed. In Section 4 brief conclusions are drawn.

\section{Real and simulated catalogs}

The catalog of peculiar velocities we consider here is the same used
in TMLM, where more details can be found. It was compiled from the ``Mark II"
data sample, which is a collection of different samples including spirals,
ellipticals and S0 galaxies. In order to reduce distance uncertainties, we
grouped the
galaxies following the rules in the original papers (Lynden--Bell \et 1988;
Faber \et 1989). Our sample finally consists of 1184 galaxies grouped in 704
objects (see Table 1 in TMLM).

To simulate the large--scale peculiar velocity field traced by optical
galaxies we performed N--body simulations of the matter distribution.
We used a particle--mesh code with $128^3$
particles on $128^3$ grid points; the simulation box is
$260~h^{-1}$ Mpc large, much larger than the typical depth of the real
galaxy sample: in fact the maximum galaxy distance is $\approx 12,000$ km
s$^{-1}$, but about 90\% of the galaxies in the sample has a distance smaller
than 6,000 km s$^{-1}$.

The initial velocity field is assumed to be Gaussian distributed with
power--spectrum $P_v(k) \propto  k^{n-2} T^2(k)$, where $T(k)$ is the CDM
transfer function $T(k) = [1 + 6.8 k + 72.0 k^{3/2} + 16.0 k^2 ]^{-1}$
(Davis \et 1985).

We ran simulations with four values of the primordial spectral index ($n=0.6,
0.8, 1, 1.2$), considering different values for the linear biasing parameter
$b$. In particular we consider $b=2, 2.5$ for $n=0.6$; $b=1, 1.5,
2, 2.5$ for $n=0.8$ and $n=1$; $b=0.5, 0.75, 1, 1.5$ for
$n=1.2$. These values are in the range allowed by COBE data, when considered
with their error bars. In the following analysis we will only plot
the results for the ``basic" models, i.e. those with the value of $b$
closest to the best fit from COBE data, when the gravitational--wave
contribution is neglected (i.e. $G(n) = 1$): these are
($n$, $b$)=(0.6,~2.5), (0.8,~1.5), (1,~1) and (1.2,~0.5). For all the
other models we report the results only in the tables for completeness.

In order to recover the velocity field from each simulation, we follow a
standard procedure (e.g. Kofman \et 1994). First, we interpolate the
mass and momentum from the particle distribution onto a cubic grid with
$128^3$ grid points, using a TSC algorithm (e.g. Hockney \& Eastwood 1981);
then, we smooth with a further Gaussian filter with width $2~h^{-1}$ Mpc,
to ensure a non zero density at every grid point. Finally, we define the
velocity at each grid point as the momentum divided by the mass density.

In Figures 1a and 1b we plot the projected particle positions and the smoothed
peculiar velocity taken from a slice of depth $4~h^{-1}$ Mpc,
for the basic models.
A first glance suggests that the dominant effect is given by the
value of $b$. The tilted models, which are characterized by high power on
large scales, are unable to fully display this feature, because of the low
evolution implied by the {\em COBE} normalization. Conversely, the blue
model, due to the long period of non--linearity occurred, appears as the one
with the largest and densest structures accompanied by coherent flows.

\bigskip
{\centerline {\bf FIGURES 1a and 1b}}
\bigskip

Our simulated catalogs are built up after choosing from each simulation 500
``observers", corresponding to grid points with features similar to those of
the Local Group (LG) (e.g. Gorski \et 1989; Davis, Strauss, \& Yahil 1991;
Strauss, Cen, \& Ostriker 1993). The requirements (slightly updated with
respect to those applied in TMLM) are the following:

\noindent
i) The peculiar velocity $v$ is in the range of the measured LG motion,
$v_{LG,obs} = 627 \pm 22$ km ${\rm s^{-1}}$ (Kogut \et 1993).

\noindent
ii) The local flow is quiet, as expressed by the the small value of the
local `shear', ${\cal S} \equiv |{\bf v} - \langle {\bf v} \rangle|/| {\bf v}|
< 0.2$, where $\langle {\bf v} \rangle$ is the average velocity of a sphere of
radius $R= 750$ km ${\rm s^{-1}}$ centered on the LG.

\noindent
iii) The density contrast in the same sphere is in the range
$-0.2 \le \delta \le 1.0$.

Radial peculiar velocities are measured by sampling the velocity field from
the LG position. We fix the axes of our reference frame by imposing that the
velocity of each simulated Local Group singles out the CMB dipole direction
($l=276\pm 3^\circ$, $b=30\pm 3^\circ$; Kogut \et 1993), with the direction
of the remaining axis randomly selected; we account for observational errors
in the direction determination by adding a Gaussian random shift
with the given
uncertainty. Next we built up our simulated catalogs by collecting, for each
of the 704 positions of the selected objects, the closest particle in the
simulation and projecting its velocity along the line of sight of the mock
LG; this method is applied in order to avoid the velocity smoothing, which
would be automatically present if the velocity field was reconstructed using
a grid (see, e.g., TMLM).

Random galaxy distance errors are then considered by perturbing each distance
and radial peculiar velocity with Gaussian noise (e.g. Dekel, Bertschinger, \&
Faber 1990): $r_{i,p} = r_i + \xi_i \Delta r_i $ and $u_{i,p} = u_i - \xi_i
\Delta r_i + \eta_i \sigma_f,$ where $\xi_i$ and $\eta_i$ are independent
standard Gaussian variables; $\Delta r_i$ is the estimated galaxy distance
error and $\sigma_f=200$ km ${\rm s^{-1}}$ is the Hubble flow noise.

\section{Statistical tests and Maximum Likelihood analysis}

In this Section we will compare the real sample with the mock catalogs using
different statistics: the Local Group constraints, the bulk flow, the Cosmic
Mach Number and the velocity correlation function.

\subsection{Local Group statistics}

The first statistic we applied to our mock catalogs is based on characterizing
the observers (Local Groups) by their velocity, local shear and local
density contrast, as previously described. We found that the ensemble of these
LG constraints changes the estimates of the other statistics: for instance, by
imposing the quietness of the local flow and allowing for a narrow range for
the local velocity we exclude grid points with large velocity.
Moreover, by considering points with a density contrast very close to the mean
avoids the choice, as origin of our simulated catalogs, of structures like
(Great) attractors, where the flow is very peculiar.

In Figure 2 we show for our basic models ($n$, $b$)=(0.6,~2.5), (0.8,~1.5),
(1,~1) and (1.2,~0.5) the probability distribution of the quantities $v$,
$\delta$ and $\cal S$: the shaded regions in the histograms show the range
allowed by the assumed LG constraints.

\bigskip
{\centerline {\bf FIGURE 2}}
\bigskip

Table 1 reports the percentage of grid points, for all the considered models,
that fulfil each constraint separately and altogether [${\cal P}(LG)$].

\bigskip
{\centerline {\bf TABLE 1}}
\bigskip

Even though the allowed range for the local shear has been restricted
with respect to TMLM (from ${\cal S}<0.5$ to ${\cal S}<0.2$), we
still find that this constraint is poorly effective: the
differences among the considered models are very small and
do not help in discriminating among them. On the contrary,
the constraints on the density and velocity of the LG turn out to
be strongly dependent on the bias parameter but almost independent of the
spectral index: higher is the value of $b$, higher is ${\cal P}(\delta)$ and
lower ${\cal P}(v)$, for all the considered $n$; the only
exception to this trend is for the most evolved model ($n$, $b$)=(1.2, ~0.5).
Finally, the velocity constraint
is the most effective one. These results generally confirm the linear analysis
performed in TMLM. The total probability ${\cal P}(LG)$ selects as best models
those in the region $(n,~b) = (0.8-1,~1.5)$.

In Figure 2, we also plot the probability distributions, for SCDM with $b=1$,
obtained within linear theory by Monte Carlo simulations in TMLM.
Note that, besides being unable to reproduce the actual skewness
of the mass distribution, the linear theory tends to overestimate both
the velocity and the shear.

\subsection{Bulk flow}

The velocity dipole or {\em bulk flow} for a galaxy catalog with $N$ objects,
with peculiar velocities ${\bf v}_i$, can be defined through a
least--squares fit (e.g. Reg\H{o}s \& Szalay 1989)
\be
v_{bulk}^\alpha = {(M^{-1})^{\alpha\beta} \sum_{i=1}^N w_i
u_i^\beta \over \sum_{i=1}^N w_i}
\ee
(summation over repeated indices is understood); $u_i^\alpha \equiv ({\bf v}_i
\cdot {\hat {\bf r}}_i) {\hat {\bf r}}_i^\alpha$ is the $\alpha$ component
($\alpha=1,2,3$) of the radial peculiar velocity and $w_i$ the weight assigned
to the $i$--th galaxy. The {\em weighted projection matrix} $M^{\alpha\beta} =
\sum_{i=1}^N w_i \hat r_i^\alpha \hat r_i^\beta/\sum_{i=1}^N w_i$
accounts for the sample geometry.

As discussed in TMLM, where different choices for the weight were considered,
the ``number weighting" scheme, $w_i=1$, always
provides a better estimate of the
true velocity dipole: for this reason we will apply in this paper this scheme
also to the other statistics considered (Cosmic Mach Number and velocity
correlation function).

In Figure 3 we plot the distribution of bulk flow amplitudes calculated from
our mock catalogs. The continuous vertical line refers to the
observed value: for our composite galaxy sample we find $v_{bulk}=306 \pm 72$
km ${\rm s^{-1}}$, with a misalignment angle $\alpha = 54^\circ \pm 13^\circ$
with respect to the direction of the CMB dipole. The plotted error bars take
into account the uncertainties due to both sparse geometry
(obtained by bootstrap resamplings of the real catalog) and distance errors
(obtained by Monte Carlo error propagation on the observed sample).
The figure suggests that the leading effect is once more determined by
the value of  $b$: higher the normalization (lower $b$) higher the
mean bulk flow and larger the spread of the distribution.

\bigskip
{\centerline {\bf FIGURE 3}}
\bigskip

In linear theory the three components of the peculiar velocity field
are independent Gaussian fields; it follows that the probability
distribution for the bulk flow of a sample is a Maxwellian,
\be
{{\cal P}(x)dx}= {32 \over {\pi^2}}
{x^2 \over {\langle x \rangle^3}}
\exp \left( -{4 \over \pi} {x^2 \over \langle x \rangle^2}\right) dx .
\ee
Suto \& Fujita (1990), using N--body simulations, found that
the bulk flow data can be fitted by a Maxwellian distribution also
in the mildly nonlinear regime.
By a Kolmogorov--Smirnov test, we checked that this curve provides a
good representation also for the mock data extracted from our simulations.
The resulting curves are also shown in Figure 3.

In Table 2, we report the probability ${\cal P}(v_{err})$ that the
simulated bulk flows have amplitude in the interval $[v_{obs} -
\sigma_{v obs},~v_{obs} + \sigma_{v obs}]$ and misalignment angle
$\alpha$ in the analogous interval. These results confirm our guess that
the spectral index plays only a limited role in this context.

\bigskip
{\centerline {\bf TABLE 2}}
\bigskip

\subsection{Cosmic Mach Number}

Another relevant statistic for the peculiar velocity field is the
{\em Cosmic Mach Number} $\cal M$, first proposed by Ostriker \& Suto
(1990). Given a galaxy sample, $\cal M$ is defined as the ratio of the
center--of--mass velocity of the sample (the bulk flow) to the one--point
velocity dispersion around this average motion.
Since the bulk flow is caused by density fluctuations on scales larger
than the sampled volume, while the velocity dispersion mostly depends
on the power on smaller scales, $\cal M$ actually measures the ratio
of large to small scale power in the velocity field.

In its first implementation the Cosmic Mach Number has been
studied in linear theory, and analytical results have been obtained based on
idealized observations. These results show that $\cal M$ is independent of
the $P(k)$ normalization, i.e. of $b$, and rule out SCDM at the $95\%$
confidence level.
In an analysis based on N--body simulations, Suto, Cen, \& Ostriker (1992)
found that the value
of $\cal M$ is well correlated with the bulk flow amplitude $v_{bulk}$, but
practically uncorrelated with the residual one--point velocity dispersion
$\sigma_v$. From this study they were able to reject SCDM at the
$99\%$ confidence level. Strauss, Cen, \& Ostriker (1993)
were the first to consider a working definition of $\cal M$,
taking into account the actual observational limitations and uncertainties.
In their analysis they tried different versions of the test applied to
three galaxy samples: a subsample of the Aaronson \et (1986,
1989) spirals, the ellipticals in Faber \et (1989) and the spiral sample
of Willick (1991). Using a procedure quite similar to the one
adopted in TMLM and in this work, they analyzed
the predictions of different cosmological models and came to the
rejection of unbiased SCDM at about the $90\%$ confidence level,
when the Aaronson \et subsample is considered.

As we did for the other statistical tests, we will use an operative
definition for $\cal M$, similar to that in Eq.(4) of Strauss,
Cen, \& Ostriker (1993) for the uniform weighting scheme, namely
\be
{\cal M} = {|v_{bulk}| \over \sqrt{3}\sigma_v},
\ee
where the bulk flow amplitude $v_{bulk}$ has been previously defined and
$\sigma_v$ is the radial part of the one point velocity dispersion in
the reference frame of the bulk motion:
\be
\sigma_v^2 = \langle (u_i - v_{bulk}^\alpha \hat r_i^\alpha)^2\rangle
= {1\over N} \sum_i (u_i -  v_{bulk}^\alpha \hat r_i^\alpha)^2 ,
\ee
with the sum extending over all the objects in the sample.
The factor $\sqrt 3$ in this expression follows from the assumption
of isotropic dispersion for the residual velocities, used in
order to estimate the full three--dimensional $\sigma_v$ from its radial
component.

For our composite sample we find ${\cal M}=0.24\pm0.06$, where,
as before, the quoted error takes into account the
uncertainties deriving from sparse geometry and distance errors.
In Figure 4 we show the Cosmic Mach Number distribution, calculated from
our mock catalogs. The shaded region indicates the observed value and its
error bar. The dashed line refers to
our best fit with a Maxwellian distribution, which once again provides a
good fit of the data, as shown by a Kolmogorov--Smirnov test (see also
Suto \& Fujita 1990; Strauss, Cen, \& Ostriker 1993).

Note that, contrary to the linear theory prediction, the Cosmic Mach Number
depends on the value of $b$, which gives in fact the dominant dependence.

\bigskip
{\centerline {\bf FIGURE 4}}
\bigskip

In Table 2, we report the probability ${\cal P}({\cal M}_{err})$ that the
simulated Cosmic Mach Number is inside the interval ${\cal M}=0.24\pm 0.06$.
Even though this statistic is the less stringent one, on its basis we can
conclude that lower values of $b$ are preferred.

It is worth to point out that our results differ from those of
Strauss, Cen, \& Ostriker (1993), mainly because the statistical approaches
are different in a number of ways.
First, we compute the Cosmic Mach Number for the whole Mark II catalog;
we expect that such a measure is more stable and representative of the
velocity field than measures performed on subsamples.
Second, we are relying on a likelihood analysis to draw our conclusions.
This means that we are not rejecting a model on the basis of
its {\em absolute} inability to reproduce the observed $\cal M$,
but rather comparing the performance of different models ``normalized''
to the performance of the maximum likelihood model.
With this warning in mind, we may try to compare our result with those
of Strauss  Cen, \& Ostriker (1993). Repeating the analysis outlined
above, but using only the sample of ``Good'' spirals of Aaronson \et, with
estimated distance less than $3000$ km ${\rm s^{-1}}$, we found that
$6\%$ to $10\%$ of the mock catalogs have a Mach Number larger than the
observed one ($0.76\pm 0.10$) for the unbiased SCDM model, corresponding to a
rejection of the model at the $90\%$ confidence level. This is in complete
agreement with Strauss Cen, \& Ostriker (1993), considering the residual
differences of the two approaches.
Rephrasing the same analysis in terms of likelihood, $15\%$ of the simulated
catalogs have a Cosmic Mach Number equal (within observational errors)
to the one observed for the same spirals. Comparing such figures with the
corresponding ones for the total sample we see that the spirals
pose a harder challenge to the unbiased SCDM model than the whole Mark II
catalog.

\subsection{Velocity correlation function}

The last statistic we consider is the {\em velocity correlation function};
following Gorski \et (1989) we define
\be
\Psi_1(r) = {\sum_{pairs(r)} {\bf u}_1 \cdot {\bf u}_2 \over
\sum_{pairs(r)} (\hat {\bf r}_1 \cdot \hat {\bf r}_2)^2},
\ee
where the sum extends over galaxy pairs separated by a distance $r$.

Figure 5 compares the velocity correlation resulting from our mock catalogs
to the observed one. We evaluated $\Psi_1(r)$ for the
real data by counting galaxy pairs in ten separation bins of $500$ km ${\rm
s^{-1}}$ up to a maximum separation of $5,000$ km ${\rm s^{-1}}$. The error
bars, estimated as for the bulk flow, take into account both the sparse
sampling of the data and the distance errors.

\bigskip
{\centerline {\bf FIGURE 5}}
\bigskip

The simulated distributions look very different in different models: high
tilt and bias shrink the distribution. The widest distribution
is for $(n,~b) = (1.2,~0.5)$.

In order to compare models with observations, we adopt, among the
different possible statistics discussed in TMLM, the linear integral of
$\Psi_1(r)$ from the origin to the maximum considered pair separation,
$R_{max}=5,000$ km ${\rm s^{-1}}$ (see also Gorski \et 1989):
\be
J_v = \int_0^{R_{max}} \Psi_1(r) \ dr.
\ee
This simple one--dimensional statistic is used to compress in one
number the information carried by the function $\Psi_1(r)$. This is
done in order to reduce the dimension of the probability space associated
to the statistic, thus ensuring a more accurate sampling.
In TMLM it was found that $J_v$ gives results in agreement with those from
the ten--dimensional sampling of $\Psi_1(r)$, showing that the compressed
statistics still retains a significant part of the original information.

For our real catalog we find $J_v/(100 {\rm ~km ~s}^{-1})^3=237.9\pm 61.5$,
where the quoted error has been estimated as for the previous statistics.
We then calculated the percentage ${\cal P} (J_{v~err})$
of the simulated catalogs whose value of $J_v$ is less than one
standard deviation different from the observed one: the results are reported
in Table 2.
In Figure 6 we show the distribution of $J_v$ calculated from our mock
catalogs. The shaded region refers to the observed value and its error bar.
As a general trend we can say that low values of $b$ are
preferred, in particular in connection with values of $n$ close to unity.

\bigskip
{\centerline {\bf FIGURE 6}}
\bigskip

\subsection{Joint Statistics}

We finally performed a Maximum Likelihood analysis to compare the statistics
obtained from different simulations. Calling $\vec C$ the random vector of the
statistics used to constrain the simulated Local Groups, $\vec C=(v_{LG},
{\cal S}, \delta)$, and $\vec S$ the vector of all the other statistics,
$\vec S=(v_{bulk}, \alpha, {\cal M}, J_v)$, the joint distribution of $\vec C$
and $\vec S$, under the condition $\vec C=\vec C_{obs}$, is
${\cal P}(\vec C_{obs},\vec S) = {\cal P}(\vec C_{obs})
{\cal P}(\vec S |\vec C_{obs})$.
For a given model $H$ (in our case for given values for $n$ and $b$),
the likelihood function reads
${\cal L}(H)={\cal P}(\vec C_{obs}|H) {\cal P}(\vec S_{obs}|\vec C_{obs},H)$.
The joint conditional likelihood ${\cal P}(\vec S_{obs}|\vec C_{obs},H)$
of $v_{bulk}$, misalignment angle $\alpha$, Cosmic Mach Number $\cal M$
and correlation integral $J_v$ has been computed by counting the number
of simulated catalogs that have, at the same time, $v_{bulk}$, $\alpha$,
$\cal M$ and $J_v$ consistent with the observed ones, within quoted error
bars. Table 2 reports, for all the considered models, the resulting values for
the joint likelihood ${\cal L}(H)$.

In order to assess merits to the different models, we can use the relative
likelihood, that is the ratio $\lambda$ of the likelihood of a model to the
maximum likelihood among them, $\lambda({\cal L})={\cal L}/{\cal L}_{MAX}$.
The quantity $- 2 \ln \lambda$ is asymptotically (i.e. for a number of
observations $N \to \infty$) distributed as a {\em Chi--square} around this
maximum, with a number of degrees of freedom equal to the number of parameters
varied in the maximization (e.g. Kendall \& Stuart 1979).
Such an approximation is also valid in the case of non--independent
observations like ours (e.g. Anderson 1971).
In our case, due to the small number of observations this is a rough
approximation, but it is nevertheless useful to provide an estimate of the
confidence levels around the maximum likelihood model. Table 2 also shows
the likelihood ratio $\lambda$ and the confidence levels (CL) for
the joint likelihood ${\cal L}(H)$.

On the basis of this analysis we can conclude that the best model
is the unbiased SCDM $(n,b)=(1,1)$. In any case we confirm the
results of TMLM on the overall flatness of ${\cal L}$
in the range $0.8 \le n \le 1$, and $1 \le b \le 1.5$. Only the models
with $b \geq 2.5$ are rejected at the $95\%$ confidence level.
The blue model, normalized to {\em COBE}, does not provide a good fit to the
data.

We also checked that removing the Cosmic Mach Number statistics from the
tests used, did not greatly change the discriminatory power of our analysis.
This is easily explained on the basis of the analysis reported below.

\subsection{Correlation among statistical tests}

The likelihood analysis we have just carried out assumes that
the velocity dipole $(v_{bulk},\alpha)$, the velocity correlation
function $\Psi_1(r)$ and the Cosmic Mach Number ${\cal M}$ of a given sample
are, in general, correlated quantities. Following this assumption
the joint likelihood is obtained by counting the catalogs that have,
{\em at the same time}, values of the statistics similar to the
observed ones. Testing such an hypothesis is of some interest because
having independent statistics would reduce the dimension of the sampled
probability distribution, which would permit higher counts and lower
statistical noise.
In this subsection we will perform a qualitative analysis of this
problem. Figure 7 shows the actual correlations between $v_{bulk}$,
$J_v$, ${\cal M}$ and $\sigma_v$ in the form of scatter plots.
For the sake of simplicity in the figure we only display
values of these statistics from the simulated catalogs of unbiased SCDM;
we found that a similar trend is provided by the other models.

\bigskip
{\centerline {\bf FIGURE 7}}
\bigskip

The plots show that the bulk flow of each sample is positively and
very well correlated to the corresponding Cosmic Mach Number;
on the contrary, there is little correlation between the latter quantity
and the residual velocity dispersion $\sigma_v$.
This feature is not new: Suto \et (1992) find the same trend in their
analysis of ${\cal M}$; their results however refer to
homogeneous and spherically symmetric velocity samples that
did not include the effects of distance errors.
We can conclude that these properties also apply when more realistic,
anisotropic and noisy velocity samples are considered.
The observed trend is easily
explained by noticing that relative departures from the respective
mean values are much smaller for $\sigma_v$ than for $v_{bulk}$; as
a consequence, it is the latter which mostly determines the value
of ${\cal M}$.
The third panel compares $v_{bulk}$ and $J_v$: it shows indeed a correlation
between the two, in the sense that higher velocities give both a higher
amplitude for the velocity dipole and a higher value for $J_v$.
The last panel shows $v_{bulk}$ vs. $\sigma_v$: the result
implies no relevant correlation between the two statistics.

{}From this analysis we can conclude that our treatment of
the data in the joint likelihood is appropriate in treating the
velocity correlation and the bulk flow as correlated quantities.
On the other hand, some improvement could be made by calculating
the probabilities for the Cosmic Mach Number from the separate
distributions of $v_{bulk}$ and $\sigma_v$, rather than from their
joint distribution. However, that would reduce by one the total
dimension of the joint probability space, which is not a big
improvement anyway.

\section{Conclusions}

In this paper we reported the results of a statistical analysis of the
large--scale velocity field in the context of $n \neq 1$ CDM models. This
extends our previous work (TMLM), based on Monte Carlo simulations within
linear theory, by the use of N--body simulations, which accounts for the
behavior of the velocity field in the non--linear regime. We considered the
values $n=0.6,~0.8,~1$ and $1.2$ for the spectral index and different values
for the bias parameter $b$. We calculated the probability to have
grid points with features similar to the Local Group; next we computed bulk
flow, Cosmic Mach Number and velocity correlation function for our mock galaxy
catalogs and compared the resulting distributions with the results of a
composite sample of 1184 galaxies, grouped in 704 objects. Using a Maximum
Likelihood method we calculated the probability of the models to reproduce
the observations, as measured by the above statistics. Our results essentially
confirm those derived from Monte Carlo simulations in TMLM, as long as the same
models are concerned.

In particular, models with high tilt ($n = 0.6$) are rejected by the
combination of the {\em COBE} results and the present analysis. The best
model is the unbiased SCDM one, $(n,~b)= (1,~1)$, but the likelihood
function is nearly flat
in the region $0.8 \le n \le 1$ and $1 \le b \le 1.5$. Note that for $n=0.8$
the values of $b$ preferred by the present analysis and consistency with
the {\em COBE} data require a negligible amount of gravitational waves.
On the other hand, tilted models with moderate bias are likely to be preferred
on the basis of a comparison with small--scale velocity dispersion data
(Davis \& Peebles 1983).

As a general result, our more accurate treatment of observational errors
shows that tilted CDM models are not excluded by the
combination of {\em COBE} data and the present analysis of
galaxy peculiar velocities. Of course,
having larger data samples, such as the ``Mark III" compilation, can help
to increase the discriminatory power of these statistical tests on the
large--scale velocity field.

\section* {Acknowledgments} We thank David Burstein for providing us with the
Mark II catalog. This work has been partially supported by MURST and by CNR
(Progetto Finalizzato: Sistemi Informatici e Calcolo Parallelo). We
acknowledge Antonio Messina for technical help and the staff and the
management of the CINECA Computer Center (Bologna) for their assistance and
for allowing the use of computational facilities.

\newpage
\large
\begin{center}
\noindent {\bf References}
\end{center}
\normalsize

\begin{trivlist}

\item[] Aaronson, M., \et 1989, ApJ, 338, 654

\item[] Aaronson, M., Bothun, G.D., Mould, J., Huchra, J., Schommer, R.A.,
\& Cornell, M.E. 1986, ApJ, 302, 536

\item[] Adams, F.C., Bond, J.R., Freese, K., Frieman, J.A,. \&
Olinto, A.V. 1993, Phys. Rev., D47, 426

\item[] Anderson, T.W. 1971, Statistical Analysis of Time Series
(New York: Wiley)

\item[] Bennett, C.L. \et 1994, ApJ, submitted

\item[] Cen, R., Gnedin, N.Y., Kofman, L.A., \& Ostriker, J.P. 1992,
ApJ, 399, L11

\item[] Crittenden, R., Bond, J.R., Davis, R., Efstathiou, G., \&
Steinhardt, P. 1993, Phys. Rev. Lett., 71, 324

\item[] Davis, M., Efstathiou, G., Frenk, C.S., \& White, S.D.M.
1985, ApJ, 292, 371

\item[] Davis, M., \& Peebles, P.J.E., 1983, ApJ, 267, 465

\item[] Davis, M., Strauss, M.A., \& Yahil, A. 1991, ApJ, 372, 394

\item[] Dekel, A., Bertschinger, E., \& Faber, S.M. 1990, ApJ, 364, 349

\item[] Faber, S.M., Wegner, G., Burstein, D., Davies, R.L.,
Dressler, A., Lynden--Bell., D., \& Terlevich, R.J. 1989, ApJS, 69, 763

\item[] Gorski, K., Davis, M., Strauss, M.A., White, S.D.M., \& Yahil,
A. 1989, ApJ, 344, 1

\item[] Hancock, S., Davies, R.D., Lasenby, A.N., Gutierrez de la Cruz,
C.M., Watson, R.A., Rebolo, R., \& Beckman, J.E. Nature, 367, 333

\item[] Hockney, R., \& Eastwood, J.W. 1981, Computer Simulation Using
Particles (New York: Mc Graw--Hill)

\item[] Kendall, M.G. \& Stuart A. 1979, The Advanced Theory of Statistics
(London: Charles Griffin \& Co.)

\item[] Kofman, L., Bertschinger, E., Gelb, J.M.,
Nusser, A., \& Dekel, A. 1994, ApJ, 420, 44

\item[] Kogut, A., \et 1993, ApJ, 419, 1

\item[] Lauer, T., \& Postman, M. 1994, preprint

\item[] Lucchin, F., Colafrancesco, S., de Gasperis G., Matarrese, S.,
Mollerach, S., Moscardini, L., \& Vittorio, N. 1994a, in preparation

\item[] Lucchin, F., \& Matarrese, S. 1985, Phys. Rev., D32, 1316

\item[] Lucchin, F., Matarrese, S., Messina, A., Moscardini, L.,
\& Tormen, G. 1994b, in preparation

\item[] Lucchin, F., Matarrese, S., \& Mollerach, S. 1992, ApJ, 401, L49

\item[] Lynden--Bell, D., Faber, S.M., Burstein, D., Davies, R.L.,
Dressler, A., Terlevich, R.J., \& Wegner, G. 1988, ApJ, , 326, 19

\item[] Maddox, S.J., Efstathiou, G., Sutherland, W.J., \& Loveday, J.
1990, MNRAS, 243, 692

\item[] Mollerach, S., Matarrese, S., \& Lucchin, F. 1994, preprint

\item[] Ostriker, J.P., \& Suto, Y. 1990, ApJ, 348, 378

\item[] Piran, T., Lecar, M., Goldwirth, D.S., da Costa, L.N., \&
Blumenthal, G.R. 1993, MNRAS, 265, 681

\item[] Reg\H{o}s, E., \& Szalay, A.S. 1989, ApJ,  345, 627

\item[] Scaramella, R., \& Vittorio, N. 1993, MNRAS, 263, l17

\item[] Smoot, G.F., \et 1992, ApJ, 396, L1

\item[] Strauss, M.A., Cen, R., \& Ostriker, J.P. 1993, ApJ, 408, 389

\item[] Suto, Y., Cen, R., \& Ostriker, J.P. 1992, ApJ, 395, 1

\item[] Suto, Y., \& Fujita, M. 1990, ApJ, 360, 7

\item[] Tormen, G., Lucchin, F., \& Matarrese, S. 1992, ApJ, 386, 1

\item[] Tormen, G., Moscardini, L., Lucchin, F., \& Matarrese, S. 1993,
ApJ, 411, 16 (TMLM)

\item[] Vittorio, N., Matarrese, S., \& Lucchin, F. 1988, ApJ, 328, 69

\item[] Willick, J. 1991, Ph.D. thesis, University of California, Berkeley

\item[] Wright, E.L., Smoot, G.F., Bennett, C.L., \& Lubin, P.M. 1994,
ApJ, submitted

\end{trivlist}

\newpage
\section*{\center Figure captions}

\noindent
{\bf Figure 1a.}
Slices with thickness $4 ~h^{-1}$ Mpc for the models $(n,~b) = (0.6,~2.5)$
and (0.8,~1.5), in the top and bottom row respectively.
Left column: projected particle positions.
Right column: projected  peculiar velocity field after smoothing by
a Gaussian filter with width $2 ~h^{-1}$ Mpc.

\vspace{0.2truecm}
\noindent
{\bf Figure 1b.}
As Figure 1a, for the models $(n,~b) = (1,~1)$ and (1.2,~0.5), in the top and
bottom row respectively.

\vspace{0.2truecm}
\noindent
{\bf Figure 2.}
Probability distribution of the peculiar velocity $v$ (top row), density
contrast $\delta$ (central row) and local `shear' $\cal S$ (bottom row),
calculated on the grid points from simulations of the models $(n,~b) =
(0.6,~2.5)$ (first column), $(n,~b) = (0.8,~1.5)$ (second column),  $(n,~b) =
(1,~1)$ (third column) and $(n,~b) = (1.2,~0.5)$ (last column). The shaded
regions show the range allowed by the different Local Group constraints.

\vspace{0.2truecm}
\noindent
{\bf Figure 3.}
Probability distribution for the absolute value of the bulk flow,
$v_{bulk}$, for the models $(n,~b) = (0.6,~2.5), ~(0.8,~1.5), ~(1,~1)$ and
(1.2,~0.5). The shaded regions refer to the one $\sigma$ range obtained
from our real catalog.

\vspace{0.2truecm}
\noindent
{\bf Figure 4.}
Probability distribution for the Cosmic Mach Number, $\cal M$, for the
models $(n,~b) = (0.6,~2.5), ~(0.8,~1.5), ~(1,~1)$ and (1.2,~0.5). The shaded
regions refer to the one $\sigma$ range obtained from our real catalog.

\vspace{0.2truecm}
\noindent
{\bf Figure 5.}
Observed velocity correlation function vs. the separation $r$ (thick solid
line with squares; error bars are one standard deviation for each bin) compared
to the probability distribution for $\Psi_1$ from the simulated catalogs for
the models $(n,~b) = (0.6,~2.5), ~(0.8,~1.5), ~(1,~1)$ and (1.2,~0.5). The
different lines refer to the $5\%$, $25\%$, $50\%$, $75\%$ and $95\%$
percentiles.

\vspace{0.2truecm}
\noindent
{\bf Figure 6.}
Probability distribution for the correlation integral $J_v$, for the
models $(n,~b) = (0.6,~2.5), ~(0.8,~1.5)$, (1,1) and (1.2,~0.5). The shaded
regions refer to the one $\sigma$ range obtained from our real catalog.

\vspace{0.2truecm}
\noindent
{\bf Figure 7.}
Correlation between different statistics for the model $(n,~b) = (1, ~1)$.
Top--left panel: the bulk flow $v_{bulk}$ vs. the Cosmic Mach Number
${\cal M}$. Top--right panel: the velocity dispersion $\sigma_v$
vs. $\cal M$. Bottom--left panel: $v_{bulk}$ vs. the correlation
integral $J_v$. Bottom--right panel: $v_{bulk}$ vs. $\sigma_v$.

\newpage
\bc
{\bf Table 1.} Local Group constraints.
\ec
\bc
\begin{tabular}{|l|l|l|l|l|l|}
\hline
\hskip0.1cm $n$ & \hskip0.1cm $b$ &
${\cal P}(v)$ & ${\cal P}(\delta$) & ${\cal P}({\cal S}$) &
${\cal P}(LG)$ \\
\hline
0.6 & ~2.0 & ~2.67 & 59.27 & 71.37 & ~~1.68 \\
0.6& ~2.5& ~0.72& 67.15& 68.91& ~~0.56\\
 & &  &  &  &  \\
0.8 & ~1.0 & ~5.97 & 37.33 & 72.49 & ~~1.62 \\
0.8&  ~1.5& ~4.66& 49.99& 68.45& ~~2.16\\
0.8 & ~2.0 & ~1.76 & 60.11 & 64.86 & ~~1.11 \\
0.8 & ~2.5 & ~0.32 & 67.88 & 61.98 & ~~0.24 \\
 & &  &  &  &  \\
1.0&  ~1.0&  ~6.41&  37.26& 73.90& ~~1.73\\
1.0 & ~1.5 & ~4.76 & 50.17 & 69.70 & ~~2.19 \\
1.0 & ~2.0 & ~1.85 & 60.32 & 65.84 & ~~1.16 \\
1.0 & ~2.5 & ~0.42 & 68.01 & 62.75 & ~~0.32 \\
 & &  &  &  &  \\
1.2& ~0.5&  ~4.64&  23.64&  69.61&  ~~0.44\\
1.2 & 0.75 & ~5.94 & 31.91 & 65.14 & ~~1.06 \\
1.2 & ~1.0 & ~5.70 & 39.34 & 61.57 & ~~1.54 \\
1.2 & ~1.5 & ~2.77 & 52.09 & 55.77 & ~~1.23 \\
\hline
\end{tabular}
\ec
\medskip
\newpage
\medskip
\bc
{\bf Table 2.} Likelihood functions.
\ec
\bc
\begin{tabular}{|l|l|l|l|l|l|l|l|}
\hline
\hskip0.1cm $n$ & \hskip0.1cm $b$ &
${\cal P}(v_{err})$ & ${\cal P}({\cal M}_{err}$) & ${\cal P}(J_{v~err}$) &
$ \hskip0.45cm {\cal L}$ & $-\ln \lambda({\cal L}) $ &
$CL({\cal L})$ \\
\hline
0.6 & 2.0 & ~~7.8 & ~~~34.6 & ~~~14.6 & ~0.030 & ~~~~0.92 & ~~60.3 \\
0.6& 2.5& ~~2.4& ~~~17.8& ~~~~9.6& ~0.003& ~~~~3.13& ~~95.6\\
 & &  &  &  &  &  &  \\
0.8 & 1.0 & ~13.4 & ~~~44.0 & ~~~23.4 & ~0.068 & ~~~~0.11 & ~~10.8 \\
0.8&1.5& ~~8.4& ~~~35.6& ~~~17.8& ~0.043& ~~~~0.57& ~~43.3\\
0.8 & 2.0 & ~~3.8 & ~~~21.6 & ~~~~7.0 & ~0.011 & ~~~~1.92 & ~~85.4 \\
0.8 & 2.5 & ~~1.2 & ~~~~9.2 & ~~~~1.8 & ~0.000 & ~~~~5.75 & ~~99.7 \\
 & &  &  &  &  &  &  \\
1.0& 1.0& ~13.2& ~~~46.6& ~~~24.4& ~0.076& ~~~~0.00& ~~~0.0\\
1.0 & 1.5 & ~10.8 & ~~~35.0 & ~~~18.6 & ~0.061 & ~~~~0.22 & ~~19.6 \\
1.0 & 2.0 & ~~5.8 & ~~~24.0 & ~~~11.4 & ~0.014 & ~~~~1.70 & ~~81.7 \\
1.0 & 2.5 & ~~3.4 & ~~~18.6 & ~~~~8.6 & ~0.003 & ~~~~3.39 & ~~96.6 \\
 & &  &  &  &  &  &  \\
1.2& 0.5& ~10.0& ~~~41.2& ~~~23.0& ~0.011& ~~~~1.97& ~~86.1\\
1.2 & 0.75 & ~13.4 &~~~43.6 & ~~~24.4 & ~0.025 & ~~~~1.10 & ~~66.7 \\
1.2 & 1.0 & ~~9.6 & ~~~38.4 & ~~~18.4 & ~0.028 & ~~~~1.01 & ~~63.6 \\
1.2 & 1.5 & ~~3.4 & ~~~21.2 & ~~~~7.4 & ~0.007 & ~~~~2.33 & ~~90.3 \\
\hline
\end{tabular}
\ec
\medskip
\end{document}